# The Planck quantum hypothesis and the Friedmannian models of flat universe

Vladimír Skalský

Faculty of Materials Science and Technology of the Slovak Technical University, 917 24 Trnava, Slovakia, skalsky@mtf.stuba.sk

**Abstract.** Only one model from an infinite number of the Friedmannian models of flat expansive isotropic and homogeneous universe satisfies the assumptions resulting from the Planck quantum hypothesis.

*Key words:* Quantum theory, cosmology

According to *the Planck quantum hypothesis* (Planck 1897a, b, c, 1898, 1899), it has a sense to think on the physical parameters of *the* observed *expansive and isotropic relativistic Universe* from the moment when its dimensions reach the values which correspond to *the Planck length* $l_P = (hG/c^3)^{1/2}$, i.e. at *the Planck time* $t_P = (hG/c^5)^{1/2}$ (Planck 1899).

The present values of (Cohen and Taylor 1999)

*the Planck constant* $\qquad h = 6.626\ 075\ 5(40) \times 10^{-34}$ J s $= 4.135\ 669\ 2(12) \times 10^{-15}$ eV s , $\qquad$ (1)

*the Newtonian constant of gravitation* $\qquad G = 6.672\ 59(85) \times 10^{-11}$ m$^3$ kg$^{-1}$ s$^{-2}$ , $\qquad$ (2)

*the speed of light in vacuum* $\qquad c = 299\ 792\ 458$ m s$^{-1}$ . $\qquad$ (3)

At present time *the Planck units* are shown with the Planck (reduced) constant $\hbar$, where: [1]

$$\hbar = \frac{h}{2\pi} . \qquad (4)$$

The present value of (Cohen and Taylor 1999)

*the Planck (reduced) constant* $\qquad \hbar = 1.054\ 572\ 66(63) \times 10^{-34}$ J s $= 6.582\ 122\ 0(20) \times 10^{-16}$ eV s . $\qquad$ (5)

The Planck units of length $l_P$ and time $t_P$ with present values $G$ (2), $c$ (3) and $\hbar$ (5) have the values:

*the Planck length* [2] $\qquad l_P = ct_P = \sqrt{\frac{\hbar G}{c^3}} = 1.616\ 04(96) \times 10^{-35}$ m , $\qquad$ (6)

*the Planck time* [3] $\qquad t_P = \frac{l_P}{c} = \sqrt{\frac{\hbar G}{c^5}} = 5.390\ 56(14) \times 10^{-44}$ s . $\qquad$ (7)

*The Friedmann general equations of isotropic and homogeneous universe dynamics* (Friedmann 1922, 1924) we can record in the form:

$$\dot{a}^2 = \frac{8\pi G a^2 \rho}{3} - kc^2 + \frac{\Lambda a^2 c^2}{3} , \qquad (8a)$$

$$2a\ddot{a} + \dot{a}^2 = -\frac{8\pi G a^2 p}{c^2} - kc^2 + \Lambda a^2 c^2 , \qquad (8b)$$

$$p = K\varepsilon , \qquad (8c)$$

where $a$ is the gauge (scale) factor, $k$ is the curvature index, $\rho$ is the mass density, $\Lambda$ is the cosmological member, $p$ is the pressure, $K$ is the state equation constant; and $\varepsilon$ is the energy density.

The Friedmannian equations (8a), (8b) and (8c) describe infinite number of *the Friedmannian models of expansive isotropic and homogeneous universe*. From these models the assumptions resulting from the Planck quantum hypothesis at the Planck time $t_P$ (7) are satisfied only by the Friedmannian model of universe in which – according to the relations (6) and (7) – for the gauge factor $a$ and the cosmological time $t$ the relation:

$$a = ct \tag{9}$$

is valid.

---

[1] At present time the presented value of the constant $h$ (1) and the constant $\hbar$ (5) do not take fully into account the relation (4). If we put as a starting point the number value of constant $h$ (1) then using the relation (4) we receive the value of $\hbar = 1.054\ 572\ 67(55) \times 10^{-34}$ J s $=$ 6.582 122 0(44) $\times 10^{-16}$ eV s. If we put as the starting point the number value of constant $\hbar$ (5) then using the relation (4) we receive the value of $h = 6.626\ 075\ 4(82) \times 10^{-34}$ J s $= 4.135.669\ 2(16) \times 10^{-15}$ eV s.

[2] At present time the presented value of the Planck length $l_P = 1.616\ 05(10) \times 10^{-35}$ m (Cohen and Taylor 1999).

[3] At present time the presented value of the Planck time $t_P = 5.390\ 56(34) \times 10^{-44}$ s (Cohen and Taylor 1999).

According to *the standard model of universe*, *the very early expansive Universe* was determined by the Friedmannian equations (8a), (8b) and (8c) with unknown values of the curvature index $k$ and the cosmological member $\Lambda$ and the value of the state equation constant $K = 1$, i.e. with

*the boundary hard state equation* $\qquad\qquad p = \varepsilon$ . $\tag{10}$

The present properties of the expansive Universe – according to the standard model of universe – are dependent on its beginning assumptions. According to the observations, the dimensions of Universe from the beginning of its expansive evolution have been increased approximately by 60 ranges, therefore – according to the standard model of universe – at the Planck time $t_P$ (7) the Universe had to have the beginning mass density $\rho_{beg} = (1 \pm 10^{-59})\rho_c$, where $\rho_c$ is the critical mass density. Therefore, although – according to the standard model of universe – we not know the present value of curvature index $k$, in the Planck time $t_P$ (7) we can regard it as $k = 0$.

In the standard model of universe is not determined unambiguously if the value of cosmological member $\Lambda$ is zero, or non-zero. Dependence of the present properties of Universe from its beginning assumptions in the standard model of universe gives restrictions not only on the value of mass density $\rho$, but and on all possible factors which would have influence on the evolution of properties of the expansive and isotropic relativistic Universe, i.e. gives restrictions and on the value of hypothetical cosmological member $\Lambda$, too. In the present cosmological literature is mostly given that in the observed Universe proportion so-called the total cosmological constant on the expansive evolution of Universe cannot be bigger than $10^{-120}$ (Abbot 1985). Therefore, in the variants of the standard universe model in most cases it is shown the value of cosmological member $\Lambda = 0$.

For the gauge factor (the overall scale of universe) $a$ and the cosmological time $t$ of the flat variant of the standard universe model – which is determined by the Friedmannian equations (8a) and (8b) with $k = 0$ and $\Lambda = 0$ and the boundary hard state equation (10) – is valid the relation (Monin *et al.* 1989):

$$a = \sqrt[3]{3ca_0^2 t} , \tag{11}$$

where $a_0$ is the chosen scale.

The relation (11) with $a_0 = l_P$ (6) and $t = t_P$ (7) gives the value of gauge factor

$$a = \sqrt[3]{3cl_P^2 t_P} = 2.330\ 74(69) \times 10^{-35}\ \text{m} = 144.225\%\ l_P . \tag{12}$$

The value of gauge factor $a$ in the relations (11), (18), (22) and (28) can be also interpreted by the relation:

$$a = vt , \tag{13}$$

where $v$ is the velocity of gauge factor $a$ increase in the cosmological time $t$.

If into the relation (13) we put $a = a$ (12) and $t = t_P$ (7) we obtain the velocity $v$ at which the gauge factor $a$ (12) increases at the time $t_P$ (7):

$$v = \frac{a\ (12)}{t_P\ (7)} = 4.323\ 75(53) \times 10^8\ \text{m s}^{-1} = 144.225\%\ c . \tag{14}$$

If in the relation (11) we put $a = a_0 = l_P$ (6) then the cosmological time

$$t = \frac{a^3}{3ca_0^2} = \frac{l_P}{3c} = 1.796\ 85(37) \times 10^{-44}\ \text{s} = 33.333\%\ t_P . \tag{15}$$

Neither the flat variant of the standard universe model, determined by the Friedmannian equations (8a), (8b) and (8c) with $k = 0$, $\Lambda = 0$ and $K = 1/3$, i.e. with

*the ultra-relativistic state equation* $\qquad\qquad p = \dfrac{1}{3}\varepsilon$ , $\tag{16}$

nor the flat variant of the standard universe model, determined by the Friedmannian equations (8a), (8b) and (8c) with $k = 0$, $\Lambda = 0$ and $K = 0$, i.e. with

*the dust state equation* $\qquad\qquad p = 0$ , $\tag{17}$

do not satisfy the assumptions, which resulted from the Planck quantum hypothesis.

For the gauge factor $a$, the chosen scale $a_0$ and the cosmological time $t$ of the flat variant of the standard universe model, determined by the Friedmannian equations (8a) and (8b) with $k = 0$ and $\Lambda = 0$ and the ultra-relativistic state equation (16), is valid the relation (Monin *et al.* 1989):

$$a = \sqrt{2ca_0 t} \;, \tag{18}$$

i.e. the gauge factor $a$ – determined by the relation (18) with $a_0 = l_P$ (6) and $t = t_P$ (7) – has the value:

$$a = \sqrt{2c l_P t_P} = 2.285\ 43(93) \times 10^{-35} \text{ m} = 141.421\% \, l_P \tag{19}$$

and – according to the relation (13) – it grows at the velocity

$$v = \frac{a\,(19)}{t_P\,(7)} = 4.239\ 70(55) \times 10^8 \text{ m s}^{-1} = 141.421\% \, c \;. \tag{20}$$

If in the relation (18) we put $a = a_0 = l_P$ (6) then the cosmological time

$$t = \frac{a^2}{2ca_0} = \frac{l_P}{2c} = 2.695\ 28(06) \times 10^{-44} \text{ s} = 50\% \, t_P \;. \tag{21}$$

For the gauge factor $a$, the chosen scale $a_0$ and the cosmological time $t$ of the flat variant of the standard universe model, which is determined by the Friedmannian equations (8a) and (8b) with $k = 0$ and $\Lambda = 0$ and the dust state equation (17), is valid the relation (Monin *et al.* 1989):

$$a = \sqrt[3]{\frac{9 a_0 c^2 t^2}{4}} \;, \tag{22}$$

i.e. the gauge factor $a$ – determined by the relation (22) with $a_0 = l_P$ (6) and $t = t_P$ (7) – has the value:

$$a = \sqrt[3]{\frac{9 l_P c^2 t_P^2}{4}} = 2.117\ 62(41) \times 10^{-35} \text{ m} = 131.037\% \, l_P \tag{23}$$

and – according to the relation (13) – it grows at the velocity

$$v = \frac{a\,(23)}{t_P\,(7)} = 3.928\ 39(25) \times 10^8 \text{ m s}^{-1} = 131.037\% \, c \;. \tag{24}$$

If in the relation (22) we put $a = a_0 = l_P$ (6) then the cosmological time

$$t = \sqrt{\frac{4 a^3}{9 a_0 c^2}} = \frac{2 l_P}{3c} = 3.593\ 70(75) \times 10^{-44} \text{ s} = 66.667\% \, t_P \;. \tag{25}$$

The Friedmannian equations (8a), (8b) and (8c) with the values $k = 0$ and $\Lambda = 0$ satisfy the assumption of relation (9) only with the value $K = -1/3$, i.e. only with (Skalský 1991)

*the zero gravitational force state equation* $\qquad p = -\dfrac{1}{3}\varepsilon \;, \tag{26}$

i.e. only in *(the Friedmannian model of) the (flat) expansive non-decelerative (isotropic and homogeneous) universe* (ENU).

Using the Friedmannian equations (8a), (8b) and (8c) with $k = 0$, $\Lambda = 0$ and $K = -1/3$, the Newtonian relations for the Euclidean homogeneous matter sphere and the famous Hubble relation (Hubble 1929) for the velocity of expanding remote objects $v$, the Hubble coefficient ("constant") $H$ and the distance of remote object $r$:

$$v = Hr \;, \tag{27}$$

result the relations between parameters of ENU (Skalský 1991):

$$a = ct = \frac{c}{H} = \frac{2Gm}{c^2} = \sqrt{\frac{3c^2}{8\pi G \rho}} \;, \tag{28}$$

where $m$ is the mass of ENU.

We can see in the relations (28) that the model of ENU is the searched by us Friedmannian model of universe that satisfies the Planckian condition of the relation (9).

More detailed analysis of the Friedmannian models of universe you can see in the paper: *The only non-contradictory model of universe* (Skalský 2000).

## Conclusions

The relations (12), (14) and (15); (19), (20) and (21); and (23), (24) and (25) show unambiguously that the flat variants of the standard universe model – determined by the Friedmannian equations (8a) and (8b) with $k = 0$ and Λ = 0 and the state equation (10), or (16), or (17) – do not satisfy the assumptions which result from the Planck quantum hypothesis at the Planck time $t_P$ (7) (Planck 1899) and do not respect nor *the relativistic principle of the boundary velocity of signal propagation c* (3) (Einstein 1905, 1916).

The only Friedmannian model of the flat universe that satisfies the assumption resulting from the Planck quantum hypothesis at the Planck time $t_P$ (7) is the ENU model – determined by the Friedmannian equations (8a), (8b) and (8c) with $k = 0$, $\Lambda = 0$ and $K = -1/3$ – which describes the expansive and isotropic relativistic universe in the Newtonian approximation (Skalský 1991, 1997, 2000).